\documentclass{ifacconf}  

\usepackage{natbib}
\usepackage{epsfig} 
\usepackage{amsmath,amssymb,amsfonts} 
\usepackage{subfigure}
\usepackage{float, color}
\usepackage{enumitem}

\usepackage{theorem}
\theoremstyle{plain} \theorembodyfont{\upshape}
\newtheorem{theorem}{Theorem}
\newtheorem{myDef}{Definition}
\newtheorem{lemma}{Lemma}
\newtheorem{myassumption}{Assumption}
\def\qed{\hfill $\Box$}
\DeclareMathOperator{\diag}{diag}
\usepackage{natbib}  

\allowdisplaybreaks[1]

\begin{document}

\thispagestyle{empty}
\pagestyle{plain}

\begin{frontmatter}

\title{
Suppressing the endemic equilibrium in SIS epidemics: A {state dependent} approach}

\thanks[footnoteinfo]{%
This work was supported in part by the Knut and Alice Wallenberg Foundation, Swedish Research Council under Grant 2016-00861 and by JSPS under Grant-in-Aid for Scientific Research Grant No. 18H01460.}

\author[First]{Yuan Wang}
\author[First]{Sebin Gracy}
\author[Second]{Hideaki Ishii}
\author[First]{and Karl Henrik Johansson}

\address[First]{
 Division of Decision and Control Systems, School of Electrical Engineering and Computer Science, KTH Royal Institute of Technology, and Digital Futures, Stockholm, Sweden. yuanwang@kth.se, gracy@kth.se, 
 kallej@kth.se}
\address[Second]{Department of Computer Science, Tokyo Institute of Technology, Yokohama, Japan. ishii@c.titech.ac.jp}


\begin{abstract}
This paper considers the susceptible-infected-susceptible (SIS) epidemic model with an underlying network structure and  focuses on the effect of social distancing to regulate the epidemic level. 
We demonstrate that if each subpopulation is informed of its infection rate and reduces interactions accordingly, the fraction of the subpopulation infected stays below half for all time instants. To this end, we first modify the basic SIS model by introducing a {state dependent} parameter 
representing the frequency of interactions between subpopulations. Thereafter, we show that for this modified SIS model, the spectral radius of a suitably-defined matrix being not greater than one causes all the agents, regardless of their initial sickness levels, to converge to the healthy state; assuming non-trivial disease spread, the spectral radius being  greater than one leads to the existence of a unique endemic equilibrium, which is also asymptotically stable. Finally, by leveraging the aforementioned results, we show that the  fraction of (sub)populations infected never exceeds half.
\end{abstract}
\end{frontmatter}

\section{Introduction}
Spreading processes such as information, diseases, and so on play an outsized role in modern societies. Notably, the ongoing COVID-19 crisis has caused disruption to our daily lives on a scale not seen in decades. Hence, spreading processes have attracted the attention of researchers {for centuries},
starting from Bernouli's seminal paper \citep{bernoulli1760essai}, 
with the key objective being to understand and 
eradicate (or, at the very least, mitigate) the spread. 
The literature abounds with relevant models, viz. susceptible-infected-recovered (SIR), susceptible-exposed-infected-recovered (SEIR), etc.
The focus of the present paper is on the  susceptible-infected-susceptible (SIS) model.

In the SIS model, an agent, which could represent either a subpopulation or an individual, is either in the susceptible or infected state. A healthy agent can get infected depending on the infection rate $\beta$, scaled by the interactions it has with its neighboring agents; in a similar manner, an infected agent recovers based on the healing rate $\gamma$. 
It is assumed  that the total number of agents is constant \citep{yorke2} and sufficiently many. The latter implies that stochastic effects can be discounted \citep{anderson1991may}.
We say that the system is in the healthy state if all the agents are healthy, or equivalently, in the disease-free equilibrium (DFE). If the epidemic remains persistent, we say that the system is in the endemic state.

Stability analysis of SIS models has been a major focus in mathematical epidemiology; see, for instance, \citep{fall2007epidemiological} and \citep{pare2018analysis} for continuous-time and discrete-time cases, respectively. Similarly, control of SIS models has also received significant attention; see, for instance \citep{torres2016sparse,watkins2016optimal}.
 We refer the interested readers to \citep{nowzari2016analysis} for an overview of these topics. 
 By leveraging the information regarding infection levels of agents, a 
 {state feedback} strategy for eradicating epidemics has been proposed \citep{pare2020data}. The strategy involves boosting the healing rates of all agents, presupposing the availability of medical resources such as vaccinations, drug administration and so on. 
In the absence of pharmaceutical intervention strategies, policymakers might have less stringent objectives. 
 
 In this paper, we approach the problem of  epidemic peak control from the viewpoint of \emph{social distancing}. Under the situation where the healthcare system is overwhelmed by the wide spread of infections, decreasing the frequency of interactions could be one of the very few effective options for mitigation; under serious conditions, its enforcement may require declarations of the state of emergency. 
 In fact, for SIR epidemics such strategies have been
 designed {previously}. The work \citep{morris2020optimal}, using the SIR model demonstrates that if social distancing is enforced effectively at {a} proper level and {an appropriate} timing, the peak of infected population can be 
reduced. In \citep{wang:ecc2021}, this model is augmented by a multi-agent system performing consensus algorithms, where the infected agents {may not behave as desired} and resilience against such {behaviors} is sought. To the best of 
our knowledge, for SIS models,  strategies for suppression of epidemics \emph{by upper bounding the proportion of {infected individuals in a subpopulation} with a specific value} are not available. We aim to address {the same} in the present paper.
  
\emph{Contributions:} 
The main contribution of this paper is to devise a control scheme for guaranteeing that the {fraction of individuals in a subpopulation} who are infected does not exceed half for all time instants. 
Our approach is as follows: First, we modify the discrete-time SIS model in \citep{pare2018analysis} by introducing a {state dependent parameter}. Then, we show that for this modified SIS model, the following properties hold:
\begin{enumerate}[label=(\roman*)]
    \item The spectral radius of a suitably-defined matrix being not greater than one guarantees convergence to the DFE; see Theorem~\ref{theorem1}.
    \item If  the spectral radius of the aforementioned matrix is greater than one, then there exists an endemic equilibrium, which has a specific characterization, and is asymptotically stable; see Theorem~\ref{theorem2}.
    \item  Finally, leveraging the results in Theorems~\ref{theorem1} and~\ref{theorem2}, we show that the fraction of infected individuals in a subpopulation never exceeds half; see Theorem~\ref{theorem3}.
\end{enumerate}

 \emph{Outline:} The rest of the paper unfolds as follows. 
The problem being investigated is formally introduced in Section~\ref{Section 2}. The main results are provided in Section~\ref{Section3}, while simulations illustrating our theoretical findings are given in Section~\ref{Section4}. Finally, a summary of the paper and some concluding remarks are provided in Section~\ref{Section5}. 

\emph{Notation:}
Let $\mathbb{R}_+$ and $\mathbb{Z}_+$ denote the sets of non-negative real numbers and integers, respectively. For any two vectors $\mathbf{a},\mathbf{b} \in \mathbb{R}^{n}$, we write {$\mathbf{a} > \mathbf{b}$} if $a_i > b_i$ for every $i \in [n]$. 
Let an eigenvalue of matrix $A$  be denoted by $\lambda(A)$. Let $\rho(A)$ 
denote the largest absolute value of an eigenvalue of matrix $A$, which is also called the spectral radius of $A$.
A diagonal matrix is denoted as $\diag(\cdot)$. We use $\mathbf{1} = [1, 1, \ldots, 1]^{T}$ and  $\mathbf{0} = [0, 0, \ldots, 0]^{T}$ to denote the vectors of all-ones and all-zeros, respectively.
Given a matrix $A$, $A \prec 0$ (resp. $ A\preccurlyeq 0 $) indicates that $A$ is negative definite (resp. negative semidefinite). 

\section{Problem Formulation}
\label{Section 2}
\subsection{Basic SIS model}
\vspace{-2ex}
Consider a network of $n$ agents, with each agent representing a subpopulation, and suppose that a virus is spreading over this network. Coming into contact with an infectious agent possibly results in an otherwise healthy agent getting infected with the virus. Such a spreading process can be represented by a directed graph $\mathcal{G}=(\mathcal{V},\mathcal{E})$, where $\mathcal V =\{1,2,\ldots,n\}$ denotes the set of agents and $\mathcal{E}$ denotes the set of interconnections between the agents. More precisely, $\mathcal E =\{(i,j) \in \mathcal{V} \times \mathcal{V} \mid a_{ij} \neq 0\}$. That is, there is a directed edge from agent $j$ to agent $i$ if a subpopulation $j$ can infect subpopulation $i$. The strength of interconnection from agent $j$ to $i$ is captured by the weight $a_{ij}>0$. Let  $\beta>0$ and $\gamma>0$ denote the infection and healing rates of each agent, respectively. 

Now, the continuous-time dynamics of each agent $i \in [n]$ can be represented as follows \citep{fall2007epidemiological}:
\begin{equation} \label{eq:ct}
\dot{x}_{i}(t) = \beta (1-x_{i}(t))  \textstyle\sum_{j=1}^{n}a_{ij}x_{j}(t)- \gamma x_{i}(t),
\end{equation}
where $x_{i}(t)$ is the infection level of agent $i$ and time $t \in \mathbb{R}_+$. 
Observe that, since the state $x_i(t)$ here denotes the fraction of the subpopulation infected at time $t$, the state values must remain in the interval $[0,1]$ and we restrict our analysis within this range for all agents.

The discrete-time version of~\eqref{eq:ct} can be obtained by applying Euler's method to \eqref{eq:ct} as in \citep{pare2018analysis}:
\begin{flalign} \label{eq:dt}
x_{i}(k+1) = & {\kern 3pt} x_{i}(k) + \Delta T\big[ \beta(1-x_{i}(k)) \textstyle\sum_{j=1}^{n}a_{ij}x_{j}(k) \nonumber \\
             & - \gamma x_{i}(k)\big],
\end{flalign}
\vspace{-3ex}

\noindent
where $\Delta T >0$ is the sampling period.

It is common to denote the basic reproduction number by $R_0 = \beta/\gamma>0$. It represents the reproduction ability, indicating how many agents an infected agent can infect on average per time step.

\vspace{-1ex}
\subsection{Problem statement}
\vspace{-2ex}
Assuming that there is a non-trivial disease spread, our goal is to devise a control scheme through social distancing such that the infection levels $x_i(k)$ of all subpopulations are bounded from above by $1/2$ for all time instants $k$.

\subsection{Modified SIS model with local policy makers}
\vspace{-2ex}

In order to achieve our goal, we modify the system in~\eqref{eq:dt} by introducing an \textit{infection reduction parameter}, denoted by $b_i(k) \in [0,1]$. This can be interpreted as a parameter provided by a local policymaker who, based on available sickness data, estimates the realtime infection level  for agent $i$ and makes preventive decisions. Such decisions, in this context, correspond to reducing the interactions with other agents in the network.
Consequently, for each agent $i$, the effective 
infection rate is reduced from $\beta$ to, at each time instant, $b_i(k)\beta$.
Hence, the dynamics in \eqref{eq:dt} 
can be written as
\begin{flalign}\label{eq-03}
x_i(k+1) =  & {\kern 3pt} x_i(k) + \Delta T \big[b_i(k)\beta(1-x_i(k)) \textstyle\sum_{j=1}^{n} a_{ij}x_j(k) \nonumber \\
            & - \gamma x_i(k)\big]. 
\end{flalign}
Note that $b_i(k) = 0$ indicates that agent $i$ removes all connections with its neighbors, while $b_i(k) = 1$ indicates all connections with neighbors are maintained as in the nominal case.

\section{Main Results}
\label{Section3}
In this section, we present a control strategy for guaranteeing that, for  $i \in [n]$, $x_i(k) \in [0, 1/2)$ for all time instants $k$. Towards this end, we set the infection reduction parameter for each agent $i$ as $b_i(k)= 1-2x_i(k)$. This indicates that each agent $i$ is asked by its local policymaker to reduce its contacts by $2x_i(k)$ {at each time instant}. 
Substituting this parameter into~\eqref{eq-03}, the dynamics for agent $i$ can be written as
\vspace{-4mm}

\begin{flalign}\label{eq-04}
    x_i(k+1) = & {\kern 3pt} x_i(k) + \Delta T \beta(1-2x_i(k))(1-x_i(k))  \nonumber \\
                 & \times \textstyle\sum_{j=1}^{n} a_{ij}x_j(k) - \Delta T \gamma x_i(k).
\end{flalign}
Since the values that $b_i(k)$ takes depend on the infection rate at {time instant $k$},
we say that it is a {state dependent parameter}.

Let $\mathbf{x}(k) = [x_1(k), x_2(k), \ldots, x_n(k)]^T$.
Then, in vector form, \eqref{eq-04} can be written as:
\begin{flalign} \label{eq-05}
    \mathbf{x}(k+1) =   & \big[ I + \Delta T \beta(I-2\mathrm{\diag}(\mathbf{x}(k)))(I-\mathrm{\diag}(\mathbf{x}(k)))A \nonumber  \\
                        &- \Delta T \gamma I\big]\mathbf{x}(k).
\end{flalign}
Observe that \eqref{eq-05} can be further rewritten as:
\begin{align} \label{eq-ex05}
\mathbf{x}(k+1) = \hat{M}(k) \mathbf{x}(k),
\end{align}
where
\begin{align}
\hat{M}(k) & =  M - B(k)A, \label{eq:Mhat}  \\
M & = I + \Delta T \beta A - \Delta T \gamma I, \label{eq-M}  \\
B(k) & = \Delta T \beta \mathrm{diag}(\mathbf{x}(k)) \big(3I - 2\mathrm{diag}(\mathbf{x}(k))\big). \label{eq-B}  
\end{align}

We need the following assumptions for our analysis.
\begin{myassumption} \citep{pare2018analysis} \label{assumption1}
The underlying graph $\mathcal{G}$ is strongly connected.
\end{myassumption}
Note that the adjacency matrix $A$ is irreducible if and only if the underlying graph $\mathcal{G}$ is strongly connected.

\begin{myassumption} \label{assumption2}
For every $i \in  [n]$, the initial state satisfies $x_i(0) \in (0, 1/2)$.
\end{myassumption}

\begin{myassumption} \label{assumption3}
$\Delta T$ is sufficiently small.
\end{myassumption}
Assumption~\ref{assumption2} ensures that {when the control action based on infection reduction starts, less than} half of the subpopulation in any agent is infected. 
Assumption~\ref{assumption3} is a technical assumption on the sampling period.

{We need the following definitions in the sequel.}
\begin{myDef}
The system \eqref{eq-05} is said to
reach the
\emph{disease free equilibrium (DFE)} if $\forall i \in [n], \lim_{k \to \infty} x_i(k) = 0$.
Also it is said to reach an
\emph{endemic equilibrium} if the states converge to a positive constant, i.e, $\forall i \in [n], \lim_{k \to \infty} x_i(k) = x^{*}_i$, where $0< x^{*}_i < 1$.
\end{myDef}

We now present our main results, whose proofs are given in the Appendix.
\begin{theorem} \label{theorem1}
Consider system~\eqref{eq-05} under Assumptions \ref{assumption1}--\ref{assumption3}.
If $\rho(M) \le 1$, then the DFE is asymptotically stable with the domain of attraction {$[0,1/2)^n$}.
\end{theorem}
Theorem~\ref{theorem1} establishes that as long as $\rho(M) \le 1$, our control scheme achieves convergence to the healthy state, {irrespective of whether the agents are initially healthy or sick. Moreover, simulations indicate that the smaller the spectral radius of $M$ is, the faster the convergence to the healthy state is; see  Fig.~\ref{fig.addnew1}}. 

It is natural to ask what the behavior of 
system~\eqref{eq-05} is when $\rho(M)>1$. We analyse the same next. 
As a first step, we introduce the following assumption. 

\begin{myassumption} \label{assumption11}
 The weights of the graph satisfy $0 \le a_{ij} <1$, \text{ and } $\textstyle\sum_{j=1}^{n} a_{ij} = 1, a_{ii} > 1/2,  \forall i \in  [n]$.
\end{myassumption}

The following lemma
establishes the relationship between $\rho(M)$ and $R_0$. 
\begin{lemma}\label{prop1}
Suppose that Assumptions \ref{assumption1} and \ref{assumption11} hold. Then $\rho(M) > 1$ if and only if $R_0 > 1$.
\end{lemma}
\emph{Proof}:
Based on Assumptions \ref{assumption1}, \ref{assumption11} and the Perron-Frobenius theorem \citep{meyer2000matrix}, it follows that $\rho(A) = 1$. 
Let $\mathbf{x}_{\lambda(A)}$ be the eigenvector for the eigenvalue $\lambda(A)$.
Then
\begin{flalign}
M\mathbf{x}_{\lambda(A)} & =  \left[ I + \Delta T \beta A - \Delta T \gamma I \right]\mathbf{x}_{\lambda(A)} \nonumber \\
                         & =  \left[ 1 + \Delta T (\beta \lambda(A) -\gamma)\right]\mathbf{x}_{\lambda(A)}. \nonumber
\end{flalign}
Then, $\lambda(M) = 1 + \Delta T (\beta \lambda(A) -\gamma)$ and thus $\rho(M) = 1 + \Delta T (\beta \rho(A) -\gamma) = 1 + \Delta T (\beta -\gamma)$. We have $\rho(M)>1$ if and only if $R_0 > 1$ (i.e., $\beta > \gamma$). 
\qed  

\begin{theorem}\label{theorem2}
Consider system \eqref{eq-05} under Assumptions \ref{assumption1}--\ref{assumption11}. If 
$\rho(M) >1$, then there exists a unique endemic equilibrium {$\overline{x}\cdot \mathbf{1} > \mathbf{0}$}, 
where
\begin{equation} \label{eq-overline_x}
\overline{x} = \frac{3R_0 - \sqrt{R_0^2 + 8R_0}}{4R_0}.
\end{equation}
Moreover, the endemic equilibrium $\overline{x} \cdot \mathbf{1}$ is asymptotically stable with the domain of attraction {$[0,1/2)^n \setminus \{\mathbf{0}\}$}.
\end{theorem}
{Theorem~\ref{theorem2} states that the reproduction number being greater than one gives rise to an endemic behavior. That is, the epidemic becomes a ``fact of life" for the community.}

We have so far shown that with $b_i(k) = 1 - 2x_i(k)$, the endemic equilibrium $\overline{x} < 1/2$ and thus $\lim_{k \to \infty}x_i(k) < 1/2$. In the following theorem, we would like to show for all $k \in \mathbb{Z}_{+}$, $x_i(k)$ is upper bounded by $1/2$. 
\begin{theorem} \label{theorem3}
Consider the system dynamics in \eqref{eq-04} 
under Assumptions~\ref{assumption1}--\ref{assumption11}.
Then,  for $i \in [n]$, 
we have $x_i(k) < 1/2$ at all times $k \in \mathbb{Z}_{+}$.
\end{theorem}
{In words, Theorem~\ref{theorem3} guarantees that the proposed control strategy  
ensures that the fraction of infected individuals in a subpopulation never exceeds half. Hence, the burden on the healthcare facilities remains more manageable.}

\section{Numerical Example}\label{Section4}
\vspace{-1ex}
\begin{figure}[t]
  \centering
      \includegraphics[width=1\linewidth]{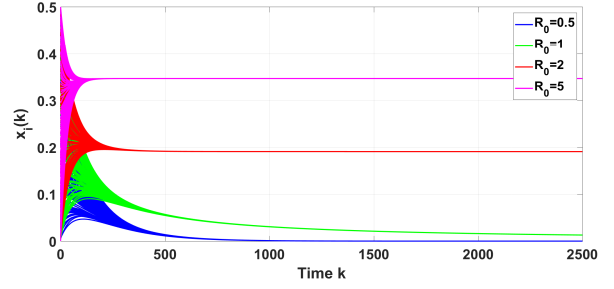}
      \caption{Time responses for $x_i(k)$ with different $R_0$}
      \label{fig.addnew1}
\end{figure}
We provide numerical examples to 
illustrate our results.
Networks with 100 agents were generated
by randomly placing
agents having the communication radius of $r =50$
in the area of 100 $\times$ 100.
For agents $i$ and $j$ that can communicate, select the weight $0 < a_{ij}<1$. The initial state $x_i(0)$ is randomly chosen from $(0,1/2)$, and $\Delta T =0.01$. We confirmed that As a result,  the conditions in Assumptions \ref{assumption1}, \ref{assumption2} and \ref{assumption11} are fulfilled.

{In this simulation, we would like to check our control strategy with different $R_0$. We test four  sets of parameters: i) $\beta =0.5, \gamma=1$, {and hence $R_0 =0.5$}; ii) $\beta =1, \gamma=1$, {and hence $R_0 =1$}; iii) $\beta =2, \gamma=1$ {and hence $R_0 =2$}; and iv) $\beta =5, \gamma=1$, {and hence $R_0 =5$}. 
Applying the policy of $b_i(k) = 1- 2x_i(k)$, the time responses for $x_i(k)$ are shown in Fig. \ref{fig.addnew1}.
{In line with the results in Theorem~\ref{theorem1}}
for the cases where $R_0=0.5$ and $R_0=1$, the states $x_i(k)$ converge to $0$ {with $R_0=0.5$ achieving exponential convergence, while for $R_0=1$ the states decay to $0$ asymptotically.}
The case when $R_0=2$ is consistent with Theorem~\ref{theorem2}, and we see  
that the endemic equilibrium {is approximately $0.19$, which indeed obeys \eqref{eq-overline_x}. Moreover,as expected, the states go to this endemic equilibrium.}  
A similar result also holds for the case $R_0=5$. Furthermore, all states are upper bounded by $0.5$, 
{consistent with the findings of} 
Theorem~\ref{theorem3}.}

 \section{Conclusion}\label{Section5}
 In this paper, we have considered a discrete-time SIS epidemic process over a strongly connected network. By leveraging the information regarding sickness levels to reduce contacts between subpopulations, we have devised a control strategy which ensures that the fraction of infected individuals in a subpopulation never exceeds half. 

\bibliography{References-Yuan}
\section*{Appendix}

We first introduce the following two lemmas for positive and non-negative matrices. 

\begin{lemma} \citep[Lemma~2]{rantzer2011distributed} \label{lemma1}
Suppose that $M$ is an irreducible non-negative matrix such that $\rho(M) <1$. Then, there exists a positive diagonal matrix $P$ such that $M^{T}PM - P \prec 0$.
\end{lemma}
\begin{lemma} \citep[Lemma~3]{pare2018analysis} \label{lemma2}
Suppose that $M$ is an irreducible non-negative matrix such that $\rho(M) =1$. Then, there exists a positive diagonal matrix $P$ such that $M^{T}PM - P \preccurlyeq 0$.
\end{lemma}

\textbf{Proof of Theorem \ref{theorem1}}: Due to Assumption \ref{assumption1}, and since $a_{ij} \ge 0$, we know that $A$ is an irreducible non-negative matrix. Therefore, from \eqref{eq-M}, it also follows that $M$ is irreducible non-negative. We separately consider the cases $\rho(M) < 1$ and $\rho(M) = 1$.

\emph{Case 1} $\rho(M) < 1$: 
From Lemma \ref{lemma1}, 
we know that there exists a
positive diagonal matrix $P_1$ such that $M^{T}P_1M - P_1 \prec 0$. 
Consider the Lyapunov candidate given by $V_1(\mathbf{x}) = \mathbf{x}^{T}P_1\mathbf{x}$ and it is immediate that $V_1(\mathbf{x}) > 0$ for every $\mathbf{x} \neq \mathbf{0}$. Let $\Delta V_1(\mathbf{x}(k)) = V_1(\mathbf{x}(k+1)) - V_1(\mathbf{x}(k))$. For $\mathbf{x}(k) \neq \mathbf{0}, k \in \mathbb{Z}_{+}$, we have
\begin{flalign}\label{eq-06}
 &\Delta V_1(\mathbf{x}(k))= \mathbf{x}^{T}(k+1)P_1\mathbf{x}(k+1) - \mathbf{x}^{T}(k)P_1\mathbf{x}(k) \nonumber \\
              &= \mathbf{x}^{T}(k) \left[\hat{M}^{T}(k)P_1\hat{M}(k) - P_1 \right] \mathbf{x}(k) \nonumber \\
              &= \mathbf{x}^{T}(k) \left[ (M - B(k)A)^{T} P_1 (M - B(k)A) - P_1 \right]\mathbf{x}(k) \nonumber \\
              &= \mathbf{x}^{T}(k) \left[ M^{T}P_1M - P_1 - M^{T}P_1B(k)A \right. \nonumber \\
              & {\kern 10pt}   \left.  - A^{T}B^{T}(k)P_1M + A^{T}B^{T}(k)P_1B(k)A \right]\mathbf{x}(k).
\end{flalign}
Since $ M^{T}P_1M - P_1$ is negative definite we have
\begin{flalign}\label{eq-07}
 \Delta V_1(\mathbf{x}(k))  & < \mathbf{x}^{T}(k) \left[ - M^{T}P_1B(k)A - A^{T}B^{T}(k)P_1M \right. \nonumber \\
              & {\kern 40pt}   \left.+ A^{T}B^{T}(k)P_1B(k)A \right]\mathbf{x}(k).
 \end{flalign}
Plugging~\eqref{eq-M} into~\eqref{eq-07}, and due to Assumption \ref{assumption3}, we have
\begin{flalign}
& \Delta V_1(\mathbf{x}(k)) \nonumber \\
&< \mathbf{x}^{T}(k) \left[  - P_1B(k)A  - \Delta T \beta A^{T}P_1B(k)A   \right. \nonumber \\
& {\kern 10pt} \left. +  \Delta T \gamma P_1B(k)A  - A^{T}B^{T}(k)P_1 - \Delta T \beta A^{T}B^{T}(k)P_1A  \right. \nonumber \\
& {\kern 10pt}  \left. + \Delta T \gamma  A^{T}B^{T}(k)P_1 + A^{T}B^{T}(k)P_1B(k)A \right]\mathbf{x}(k) \nonumber \\
& = \mathbf{x}^{T}(k) \left[ (\Delta T \gamma - 1) P_1B(k)A + (\Delta T \gamma - 1) A^{T}B^{T}(k)P_1  \right. \nonumber \\
& {\kern 10pt} - \Delta T \beta A^{T}P_1B(k)A  -  \Delta T \gamma  A^{T}B^{T}(k)P_1A \nonumber \\
& {\kern 10pt} \left. + A^{T}B^{T}(k)P_1B(k)A \right]\mathbf{x}(k) \nonumber \\
& < \mathbf{x}^{T}(k) \left[ - \Delta T \beta A^{T}P_1B(k)A  - \Delta T \beta A^{T}B^{T}(k)P_1A \right. \nonumber \\
& {\kern 10pt}   \left.+ A^{T}B^{T}(k)P_1B(k)A \right]\mathbf{x}(k) \label{eq:strict:ineq1} \\
& = \mathbf{x}^{T}(k)A^{T} \left[ - \Delta T \beta P_1B(k)  - \Delta T \beta B^{T}(k)P_1 \right. \nonumber \\
& {\kern 10pt}   \left.+ B^{T}(k)P_1B(k) \right]A\mathbf{x}(k). \label{eq-08}
\end{flalign}
Note the inequality in \eqref{eq:strict:ineq1} holds since $P_1$ and $B(k)$ are both positive diagonal matrices and $A$ is a non-negative matrix. The term $\Delta T \gamma -1$ is negative, due to Assumption \ref{assumption3}. 
Since $B(k)$ and $P_1$ are diagonal and $P_1B(k) = B(k)P_1$, from~\eqref{eq-08}, we have
\begin{flalign}\label{eq-09}
 &\Delta V_1(\mathbf{x}(k)) 
               < \mathbf{x}^{T}(k)A^{T} \left( -2\Delta T \beta I + B(k) \right)P_1B(k)A\mathbf{x}(k).
\end{flalign}
Next we consider the matrix $ \bar B(k) = -2\Delta T \beta I + B(k)$:
\begin{flalign}\label{eq-10}
\bar B(k) & = -2\Delta T \beta I + B(k) \nonumber \\
          & = -2\Delta T \beta I + \Delta T \beta \mathrm{\diag}(\mathbf{x}(k))(3I - 2\mathrm{\diag}(\mathbf{x}(k))) \nonumber \\
          & = \Delta T \beta \big[ \mathrm{\diag}(\mathbf{x}(k))(3I - 2\mathrm{\diag}(\mathbf{x}(k))) -2I  \big].
\end{flalign}
Clearly, $\bar B(k)$ is a diagonal matrix and its $i$th element is
$[\bar B(k)]_{ii} = \Delta T \beta \left( -2x_i^2(k) + 3x_i(k) -2\right)$. 
Since $0\le x_i(k)\le 1$, we know that $[\bar B(k)]_{ii} \le (-7/8)\Delta T \beta < 0$, which indicates that $\bar B(k)$ is negative definite. Moreover, since $A, A^{T}, P_1$ and $B(k)$ are all non-negative matrices, we conclude that $\Delta V_1(\mathbf{x}(k)) < 0$ from \eqref{eq-09}. Therefore, from \citep{vidyasagar2002nonlinear}, the system converges asymptotically to the DFE for this case.

\emph{Case 2} $\rho(M) = 1$: Due to Lemma \ref{lemma2}, the condition $\rho(M) = 1$ guarantees the existence of a positive diagonal matrix $P_2$ such that $M^{T}P_2M - P_2 \preccurlyeq 0$. 
Consider the Lyapunov candidate given by  $V_2(\mathbf{x}(k)) = \mathbf{x}^{T}(k)P_2\mathbf{x}(k)$.  
The rest of the proof is quite similar to 
the case of $\rho(M) <1$, and, hence, we arrive at $\Delta V_2(\mathbf{x}(k)) < 0$. Therefore, from \citep{vidyasagar2002nonlinear} it follows that the system converges asymptotically to the DFE for this case as well.
\qed

\textbf{Proof of Theorem \ref{theorem2}}: The proof is inspired by \cite{fall2007epidemiological} and \cite{liu2020stability}. It consists of two steps: we first establish the existence and uniqueness of the endemic equilibrium. Subsequently, we establish, for all non-zero initial conditions, asymptotic convergence to the said equilibrium.

\textit{Step 1: Existence/Uniqueness of the endemic equilibrium}\\
By \eqref{eq-05}, an equilibrium $\mathbf{x}^* = [x_1^*, x_2^*, \ldots, x_n^*]^T$ satisfies
\begin{flalign}
    \mathbf{x}^* =  {\kern 3pt} & \big[ I + \Delta T \beta(I-2\mathrm{\diag}(\mathbf{x}^*))(I-\mathrm{\diag}(\mathbf{x}^*))A \nonumber \\
     & - \Delta T \gamma I\big]\mathbf{x}^*. \nonumber
\end{flalign}
Hence, it follows that
\begin{flalign}
\left( I - R_0 A  \right) \mathbf{x}^* = {\kern 3pt} & 2 R_0 \mathrm{\diag}(\mathbf{x}^*) \mathrm{\diag}(\mathbf{x}^*) A \mathbf{x}^* \nonumber \\ 
& -3R_0\mathrm{\diag}(\mathbf{x}^*) A \mathbf{x}^*. \nonumber
\end{flalign}
Furthermore, we have
\begin{flalign}
& \big[ I + \mathrm{\diag}(3R_0A\mathbf{x}^*) -  \mathrm{\diag}(2R_0A\mathbf{x}^*)\mathrm{\diag}(\mathbf{x}^*)\big]\mathbf{x}^*
 = R_0 A \mathbf{x}^*,
\nonumber
\end{flalign}
Define $H(\mathbf{x}^*) = I+ \mathrm{\diag}(3R_0A\mathbf{x}^*) -  \mathrm{\diag}(2R_0A\mathbf{x}^*)\mathrm{\diag}(\mathbf{x}^*)$.
It can be immediately seen that $H(\mathbf{x}^*)$ is a positive diagonal matrix with $[H(\mathbf{x}^*)]_{ii} \ge 1$, and as a consequence $H^{-1}(\mathbf{x}^*)$ exists. Thus we have
\begin{flalign} \label{eq-12}
\mathbf{x}^*=H^{-1}(\mathbf{x}^*)R_0A \mathbf{x}^*
\end{flalign}
By assumption, $\rho(M)>1$. Hence, due to Lemma~\ref{prop1}, it follows that $R_0 > 1$, and, hence we can choose a small $\varepsilon$ satisfying $0< \varepsilon < \frac{3R_0 - \sqrt{R_0^2 + 8R_0}}{4R_0}$. It then holds $0 < 1+3R_0\varepsilon -2R_0\varepsilon^2 < R_0$
and thus $\varepsilon < \frac{R_0\varepsilon}{1+3R_0\varepsilon-2R_0\varepsilon^2}$.\\
Furthermore due to Assumption \ref{assumption11}, 
it follows that $R_0A\mathbf{1}=R_0\mathbf{1}$.
Then, for $i \in [n]$, we have
\[
\varepsilon < \frac{[R_0A\mathbf{1}]_i \varepsilon}{1+3[R_0A\mathbf{1}]_i\varepsilon-2[R_0A\mathbf{1}]_i\varepsilon^2}.
\]
Hence, it follows that
\[
\varepsilon \cdot\mathbf{1} < H^{-1}(\varepsilon \cdot\mathbf{1})R_0A (\varepsilon \cdot \mathbf{1}).
\]
Similarly, by taking $\mu$ satisfying $\frac{3R_0 - \sqrt{R_0^2 + 8R_0}}{4R_0} < \mu < 1$, we have
\[
\mu \cdot\mathbf{1} > H^{-1}(\mu \cdot\mathbf{1})R_0A (\mu \cdot \mathbf{1}).
\]
It is clear that $\overline{x} = \frac{3R_0 - \sqrt{R_0^2 + 8R_0}}{4R_0}$ satisfies
\[
\overline{x}\cdot \mathbf{1} = H^{-1}(\overline{x}\cdot \mathbf{1})R_0A (\overline{x}\cdot \mathbf{1}).
\]

We prove uniqueness by a contradiction argument. Suppose that there is another endemic equilibrium $\overline{\mathbf{x}}^* = [\overline{x}_1^*, \overline{x}_2^*, \ldots, \overline{x}_n^*]$. Let
$\zeta = \max_{i \in [n]}{\frac{\overline{x}_i^*}{\overline{x}}}$.
We would like to show that $\zeta \le 1$. By way of contradiction, assume that $\zeta >1$. 
This implies that
$\overline{\mathbf{x}}^* \le \zeta \overline{x} \cdot \mathbf{1}$ and there exists an $i_0$ such that $\overline{x}_{i_0}^* = \zeta \overline{x}$. We note that $\overline{x}_{i_0}^* \le 1$ so that $1< \zeta \le 1/{\overline{x}}$.
Define a map $f(x_i): \mathbb{R}_+ \rightarrow \mathbb{R}_+$ such that $f(x_i)=[H^{-1}(\mathbf{x}^*)R_0A \mathbf{x}^*]_i$. 
Then, for the aforementioned node $i_0$, based on \eqref{eq-12} 
and {since} $f(\overline{x}^*_{i_0}) = f(\zeta \overline{x})$, we have
\begin{flalign} \label{eq-13}
\overline{x}_{i_0}^*
& =
\frac{[R_0A\overline{\mathbf{x}}^*]_{i_0}}{1+3[R_0A\overline{\mathbf{x}}^*]_{i_0}-2[R_0A\mathrm{\diag}(\overline{\mathbf{x}}^*)\overline{\mathbf{x}}^*]_{i_0}} \nonumber \\
& = \frac{[R_0A\zeta \overline{x} \cdot \mathbf{1}]_{i_0}}{1+3[R_0A\zeta \overline{x} \cdot \mathbf{1}]_{i_0}-2[R_0A\mathrm{\diag}(\zeta \overline{x} \cdot \mathbf{1})\zeta \overline{x} \cdot \mathbf{1}]_{i_0}} \nonumber \\
& = \frac{\zeta[R_0A \overline{x} \cdot \mathbf{1}]_{i_0}}{1+3\zeta[R_0A\overline{x} \cdot \mathbf{1}]_{i_0}-2\zeta^2[R_0A\mathrm{\diag}( \overline{x} \cdot \mathbf{1}) \overline{x} \cdot \mathbf{1}]_{i_0}} \nonumber \\
& = \frac{ R_0\zeta \overline{x}}{1+3 R_0\zeta \overline{x} - 2R_0\zeta^2  \overline{x}^2}.
\end{flalign}
Let
\begin{flalign}
g(\zeta) & = \left[1+3 R_0\zeta\overline{x} - 2R_0(\zeta \overline{x})^2\right] - \left[1+3 R_0 \overline{x} - 2R_0\overline{x}^2\right] \nonumber \\
         & = 3R_0\overline{x} (\zeta-1) - 2R_0\overline{x}^2(\zeta-1)(\zeta+1) \nonumber \\
         & = R_0\overline{x} (\zeta-1) (3 -2\overline{x}\zeta - 2\overline{x}). \nonumber
\end{flalign}
Since $\zeta \overline{x}\le 1$ and $\zeta > 1$, we have
\begin{flalign}
g(\zeta) & \ge R_0\overline{x} (\zeta-1) (1 - 2\overline{x}) >0. \nonumber
\end{flalign}
Thus, from \eqref{eq-13}, we have
\begin{flalign} \label{eq-14}
\overline{x}_{i_0}^* < \frac{\zeta R_0 \overline{x}}{1+3 R_0 (\overline{x}) - 2R_0(\overline{x})^2} = \zeta\overline{x} = \overline{x}_{i_0}^*.
\end{flalign}
Hence, we obtain a contradiction of the assumption that $\zeta >1$, thus implying $\zeta \le 1$. 
{Therefore,} if there exists another equilibrium $\overline{\mathbf{x}}^*$, it must satisfy $\overline{\mathbf{x}}^* \le \overline{x} \cdot \mathbf{1}$. By exchanging the roles of$\text{ }\overline{\mathbf{x}}^*$ and $\overline{x} \cdot \mathbf{1}$, by a similar analysis as before, we obtain $\overline{\mathbf{x}}^* \ge \overline{x} \cdot \mathbf{1}$. This implies $\overline{\mathbf{x}}^* = \overline{x} \cdot \mathbf{1}$, 
thus concluding the proof of uniqueness.

\textit{Step 2: Stability of the endemic equilibrium}\\
First, note that any equilibrium $\overline{x}$ of~\eqref{eq-04} satisfies:
\begin{flalign}
    \overline{x} = & {\kern 3pt} \overline{x} + \Delta T \big(\beta(1-2\overline{x})(1-\overline{x}) \cdot \textstyle \sum_{j=1}^{n} a_{ij}\overline{x} - \gamma \overline{x}\big). \nonumber
\end{flalign}
Since, by Assumption~\ref{assumption11}, $\textstyle\sum_{j=1}^{n} a_{ij} =1$, and since $\overline{x}>0$, 
\begin{flalign} \label{eq-x}
\beta(1-2\overline{x})(1-\overline{x})  - \gamma  =0.
\end{flalign}
Let, for all $i \in [n]$, 
$y_i(k) = x_i(k) - \overline{x}$ 
and $\Delta y_i(k) = y_i(k+1) - y_i(k)$. Substituting $x_i(k) = y_i(k) + \overline{x}$ into \eqref{eq-04} yields 
\begin{flalign}
\Delta y_i(k)   & = \Delta T \beta \left( 1 - 2\overline{x}- 2y_i(k) \right)\left( 1 - \overline{x}- y_i(k) \right) \nonumber \\
                    & {\kern 10pt} \times \textstyle\sum_{j=1}^{n}a_{ij}\left( y_j(k) + \overline{x} \right) - \Delta T \gamma \overline{x} -  \Delta T \gamma y_i(k) \nonumber \\
                    & {\kern -10pt}= -\Delta T \gamma y_i(k) + \Delta T \beta \left( 1 - 2\overline{x} \right)\left( 1 - \overline{x} \right) \textstyle\sum_{j=1}^{n}a_{ij}y_j(k)  \nonumber \\
                    &  + \Delta T \beta \left(  2y_i^2(k) + (4\overline{x} - 3)y_i(k) \right)\textstyle\sum_{j=1}^{n}a_{ij}x_j(k)  \nonumber \\
                    &  + \Delta T \left(  \beta (1 - 2\overline{x})(1 - \overline{x}) -  \gamma \right)\overline{x}.  \nonumber
\end{flalign}
From \eqref{eq-x}, 
$\Delta T \left(  \beta (1 - 2\overline{x})(1 - \overline{x}) -  \gamma \right)\overline{x} = 0$, and {hence} 
\begin{flalign} \label{eq-15}
\Delta y_i(k)  &= \Delta T \beta \left( 1 - 2\overline{x} \right)\left( 1 - \overline{x} \right)  \textstyle\sum_{j=1}^{n}a_{ij}y_j(k) -\Delta T \gamma y_i(k) \nonumber \\
                    & {\kern -20pt} + \Delta T \beta \left(  2y_i^2(k) + (4\overline{x} - 3)y_i(k) \right)\textstyle\sum_{j=1}^{n}a_{ij}x_j(k).
\end{flalign}
Since $\overline{x} >0$, $\frac{1}{\overline{x}}$ exists.
From \eqref{eq-15}, we have the following:
\begin{flalign} \label{eq-16}
y_i(k+1) & = \left[ (1- \Delta T \gamma) + \Delta T \beta(4\overline{x} - 3)\overline{x} \right]  y_i(k)  \nonumber \\
&  + 2\Delta T \beta\overline{x}  y_i^2(k) + \Delta T \beta \big[ \left( 1 - 2\overline{x} \right)\left( 1 - \overline{x} \right)  \nonumber \\
&  + (4\overline{x} - 3) y_i(k) + 2y_i^2(k)\big] \textstyle\sum_{j=1}^{n}a_{ij} y_j(k).
\end{flalign}
{Rewriting~\eqref{eq-x} in terms of $\beta$, and plugging it into~\eqref{eq-16} yields: }
\begin{flalign} 
y_i(k+1) & = \left[ (1- \Delta T \gamma) + \Delta T \frac{\gamma (4\overline{x} - 3) \overline{x}}{(1-2\overline{x})(1-\overline{x})} \right] \nonumber \\
& {\kern 10pt} \times y_i(k) + \Delta T \frac{2\gamma \overline{x}}{(1-2\overline{x})(1-\overline{x})} \times y_i^2(k)  \nonumber \\
& {\kern -30pt} + \Delta T \beta \left[ \left( 1 - 2\overline{x} \right)\left( 1 - \overline{x} \right) + (4\overline{x} - 3) y_i(k) + 2y_i^2(k)\right] \nonumber \\
& {\kern 10pt} \times \textstyle\sum_{j=1}^{n}a_{ij} y_j(k) \nonumber \\
& = \left[ 1 - \Delta T\gamma \cdot \frac{1- 2\overline{x}^2}{(1-2\overline{x})(1-\overline{x})} \right]\times y_i(k) \nonumber \\
& {\kern 10pt}  + \Delta T \gamma \cdot \frac{2\overline{x}}{(1-2\overline{x})(1-\overline{x})} \times y_i^2(k)  \nonumber \\
& {\kern 10pt} + \Delta T \beta \left[ 1- 2(y_i(k)+\overline{x})\right] \left[1-(y_i(k)+ \overline{x})\right] \nonumber \\
& {\kern 10pt} \times \textstyle\sum_{j=1}^{n}a_{ij} y_j(k). \nonumber \\
& = \left[ 1 - \Delta T\gamma \cdot \frac{1- 2\overline{x}\cdot (\overline{x} + y_i(k))}{(1-2\overline{x})(1-\overline{x})} \right]\times y_i(k) \nonumber \\
& {\kern 10pt} + \Delta T \beta \left[ 1- 2(y_i(k)+\overline{x})\right] \left[1-(y_i(k)+ \overline{x})\right] \nonumber \\
& {\kern 10pt} \times \textstyle\sum_{j=1}^{n}a_{ij} y_j(k). \label{eq-new17} \\
& = \left[ 1 - \Delta T\gamma \cdot \frac{1- 2\overline{x}x_i(k)}{(1-2\overline{x})(1-\overline{x})} \right]\times y_i(k) \nonumber \\
&  + \Delta T \beta \left( 1- 2x_i(k)\right) \left(1-x_i(k)\right) \textstyle\sum_{j=1}^{n}a_{ij} y_j(k). \label{eq-new18}
\end{flalign}
Note that \eqref{eq-new18}
holds since $x_i(k) = y_i(k)+ \overline{x}$. 
Let $\mathbf{y}(k) = [y_1(k), y_2(k), \ldots, y_n(k)]^T$ and we write \eqref{eq-new18} in matrix form {so as to obtain}
$\mathbf{y}(k+1) = \Phi(k) \mathbf{y}(k)$,
\vspace{-2mm}
\begin{flalign} 
\text{where } \Phi(k) =&  I - \Delta T \gamma \text{diag}\left( \frac{1}{(1-2\overline{x})(1-\overline{x})} \right)  \nonumber \\
& + \Delta T \gamma \text{diag}\left( \frac{2\overline{x}}{(1-2\overline{x})(1-\overline{x})} \right) \text{diag}\left(\mathbf{x}(k)\right) \nonumber \\
& + \Delta T \beta \left(I - 2\text{diag}\left(\mathbf{x}(k)\right) \right)\left(I - \text{diag}\left(\mathbf{x}(k)\right) \right)A \label{eq-Phi}
\end{flalign}
Construct a matrix $D$ such that
\[
D = I - \Delta T \gamma \text{diag} \left(  \frac{1}{(1-2\overline{x})(1-\overline{x})} \right) + \Delta T \beta A.
\]
We can check that $[D]_{ij} = \Delta T \beta a_{ij}, \forall i \neq j$ and 
\begin{flalign}
[D]_{ii} & = 1-  \Delta T \gamma \frac{1}{(1-2\overline{x})(1-\overline{x})} + \Delta T \beta a_{ii} \nonumber \\
& = 1 - \Delta T \beta + \Delta T \beta a_{ii} \nonumber \\
& = 1 - \Delta T \beta \textstyle\sum_{j \neq i}a_{ij}. \label{eq-dii}
\end{flalign}
Thus we have $D\mathbf{1} = \mathbf{1}$. Since $\Delta T$ is sufficiently small and for every $i\in [n]$, $\Delta T \beta \textstyle\sum_{j \neq i}a_{ij}$ is sufficiently small as well. From \eqref{eq-dii}, we have $[D]_{ii} > 0$. Then $D$ is a non-negative irreducible matrix and moreover, $\mathbf{1}>\mathbf{0}$. Hence, we have $\rho(D) =1$. In addition, based on the Perron-Frobenius Theorem for irreducible nonnegative matrices (Theorem 2.7 \citep{varga2000matrix}), we can also find a left vector $\mathbf{v}^T > \mathbf{0}^T$ such that $\mathbf{v}^T D = \mathbf{v}^T$.

Construct an auxiliary system as follows
\begin{align} \label{system_aux}
\mathbf{z}(k+1) = \Phi(k) \mathbf{z}(k),
\end{align}
where $\mathbf{z}(0) = |\mathbf{y}(0)|$. Since $\Phi(k)$ is a non-negative matrix, we have $\mathbf{z}(k) \ge \mathbf{0}$ and $ - \mathbf{z}(k) \le \mathbf{y}(k) \le \mathbf{z}(k), \forall k \in \mathbb{Z}_+$. Therefore,    $\mathbf{y}(k)$ is asymptotically stable if the origin is asymptotically stable for $\mathbf{z}(k)$. Consider Lyapunov candidate $V(k) = \mathbf{v}^T \mathbf{z}(k)$, and it can be readily seen that $V(k) \ge 0$ since $\mathbf{v}^T > \mathbf{0}^T, \mathbf{z}(k) \ge \mathbf{0}$. We note that $V(k) = 0$ only if $\mathbf{z}(k) = \mathbf{0}$. Therefore, we have
\begin{flalign}
V(k+1) - V(k) &=  \mathbf{v}^T (\mathbf{z}(k+1) -\mathbf{z}(k)) \nonumber \\
& =  \mathbf{v}^T (\Phi(k) - I) \mathbf{z}(k) \nonumber \\
& =  \mathbf{v}^T (\Phi(k) - D) \mathbf{z}(k). \label{eq-deltaV}
\end{flalign}
\begin{flalign}
\Phi(k) - D = & {\kern 3pt} \Delta T \gamma \text{diag}\left( \frac{2\overline{x}}{(1-2\overline{x})(1-\overline{x})} \right) \text{diag}\left(\mathbf{x}(k)\right) \nonumber \\
& + \Delta T \beta \text{diag}(\mathbf{x}(k))(2\text{diag}(\mathbf{x}(k)) - 3I)A. \label{eq-phi}
\end{flalign}
We note that $[\Phi(k) - D]_{ij} = \Delta T \beta_i(2x_i(k)-3)a_{ij} < 0, i \neq j$. 
{Observe that}
\begin{flalign}
[\Phi(k) - D]_{ii}  
&=  {\kern 3pt} \Delta T x_i(k) \Big[ \frac{2\gamma\overline{x}}{(1-2\overline{x})(1-\overline{x})} \nonumber \\
& {\kern 15pt}  + \beta(2x_i(k)-3)a_{ii}   \Big] \nonumber \\
&  = \Delta T x_i(k) \left[ 2\beta\overline{x} + \beta(2x_i(k)-3)a_{ii} \right] \label{eq-sta01} \\
& < \Delta T \beta x_i(k) \left[ 1 -2a_{ii} \right] <0 \label{eq-sta02}
\end{flalign}
We note that \eqref{eq-sta01} is based on \eqref{eq-x}. 
Moreover, based on $\overline{x}_i < 1/2, \forall i$, $x_i(k) < 1/2$, $a_{ii} > 1/2$ and Assumption \ref{assumption11}, we obtain \eqref{eq-sta02}. 
{Thus,} $\Phi(k) - D$ is a matrix in which every element is negative. {Then} we have
$V(k+1) - V(k) \le 0$ from \eqref{eq-deltaV}, {and} the equality {holds} if and only if $\mathbf{z}(k) =0$.
Thus, from \citep{vidyasagar2002nonlinear}, it follows that the  auxiliary system \eqref{system_aux} converges asymptotically to the origin. Thus, the system \eqref{eq-04} converges asymptotically to the endemic equilibrium. 
\qed

\textbf{Proof of Theorem \ref{theorem3}}: 
We first consider the  case where $R_0 \le 1$. Based on Assumptions~\ref{assumption1}--\ref{assumption11}  
and the results in Lemma~\ref{prop1} and Theorem~ \ref{theorem1}, for an initial state $0 < x_i(0) < {1}/{2}$, the state $x_i(k)$ asymptotically converges to $0$.
Define 
$x_{\max}(k) = \max_{i \in [n]} \{ x_i(k)\}$. Then, based on \eqref{eq-04}, we have
\begin{flalign}
    & x_i(k+1) =   x_i(k) + \Delta T \big(\beta(1-2x_i(k))(1-x_i(k))  \nonumber \\
                 & {\kern 50pt}\times \textstyle\sum_{j=1}^{n} a_{ij}x_j(k) - \gamma x_i(k)\big) \nonumber \\
            &\le    x_i(k) + \Delta T \big[\beta(1-2x_i(k))(1-x_i(k))x_{\max}(k) 
                   - \gamma x_i(k)\big]  \label{eq-ex5} \\
            &<   x_i(k) + \Delta T \big[\gamma(1-x_i(k))  
                   x_{\max}(k) - \gamma x_i(k)\big]  \label{eq-ex6} \\
            &\le   (1-\Delta T\gamma)x_{\max}(k) +\Delta T \gamma(1-x_i(k))x_{\max}(k) \label{eq-ex7} \\
            &<   x_{\max}(k). \label{eq-ex4}
\end{flalign}
The inequality \eqref{eq-ex5} is 
due to $x_j(k) \le x_{\max}(k)$ and 
Assumption \ref{assumption11}, whereas  inequality~\eqref{eq-ex6} follows due to the fact that the assumption $R_0\leq 1$ implies $\beta \le \gamma$ and, due to $1-2x_i(k)<1$. Inequality~\eqref{eq-ex7} follows due to $x_i(k) \le x_{\max}(k)$ while 
inequality~\eqref{eq-ex4} is immediate. 
Thus, for every $i \in [n]$ and $x_i(k) \neq 0$, we have $x_i(k+1) < x_{\max}(k)< x_{\max}(0) < {1}/{2}$.

Next, we consider the case where $\rho(M) >1$. Suppose that $\rho(M) >1$, then.  
since system~\eqref{eq-04} satisfies Assumptions~\ref{assumption1}-\ref{assumption11}, from Theorem~\ref{theorem2}, we know that there exists endemic equilibrium $\overline{\mathbf{x}}$ such that for every $i \in [n]$, $0<\overline{x}_{i} <1/2$. 

We note that $R_0>1$ if $\rho(M) >1$. In the following analysis, we would like to consider the following two cases:

\emph{Case 1}: The local states that satisfies  
\[
\frac{1}{2}\left( 1- \frac{1}{R_0} \right) < x_i(k) < \frac{1}{2}. 
\]
Based on \eqref{eq-03}, we have the following inequalities:
\begin{flalign}
     &x_i(k+1)  \le x_i(k) + \Delta T \gamma  \big((1-2x_i(k))(1-x_i(k)) \nonumber \\
                 & {\kern 50pt}\times x_{\max}(k)R_0 - x_i(k)\big) \label{eq-add5} \\
    & < x_i(k) + \Delta T \gamma \big((1-x_i(k)) x_{\max}(k) - x_i(k)\big) \label{eq-add3} \\
    & < x_{\max}(k). \label{eq-add4}
\end{flalign}
We note that \eqref{eq-add3} hold since $\frac{1}{2}\left( 1- \frac{1}{R_0} \right)  < x_i(k) < 1/2$, then we have $(1-2x_i(k))R_0 < 1$.
Inequality~\eqref{eq-add4} follows from the same line of reasoning as inequality~\eqref{eq-ex5}.

\emph{Case 2}: The local states that satisfies  
\[
0 < x_i(k) < \frac{1}{2}\left( 1- \frac{1}{R_0} \right). 
\]

Since the initial states satisfies $0 < x_i(0)< 1/2$, we have that $x_i(k)< 1/2, \forall i \in [n]$. The monotonicity is unclear in this case. If $x_i(k)$ does not exceed the interval, it is clear that we have $0<x_i(k)<1/2$, $\forall k \in \mathbb{Z}_{+}$. 
Otherwise, from \eqref{eq-04}, we have
\[
x_i(k+1) \le x_i(k) + \Delta T \beta \textstyle \sum_{j=1}^{n}a_{ij}.
\]
Based on Assumption \ref{assumption3}, $\Delta T$ is sufficiently small such that the increment $\Delta T \beta \textstyle \sum_{j=1}^{n}a_{ij} $ is sufficiently small as well. Then there must exists some time $k' > k$ such that $x_i(k')$ satisfies Case 1.

Thus, for all agents with initial local states that satisfies Assumption \ref{assumption2}, we always have $x_i(k) < {1}/{2}$ for all $k \in \mathbb{Z}_{+}$. 

\qed
\end{document}